# Atlantis, Carli and Bailly, and a short discussion about 'demonstration' in cultural astronomy

Elio Antonello
*INAF-Astronomical Observatory of Brera*
*Italian Society for Archaeoastronomy*
elio.antonello@brera.inaf.it

**Abstract.** Some European scholars in the second half of 18[th] century discussed the myth of Atlantis taking into account the studies of the evolution of the Earth and of the beginning and diffusion of civilization, in particular the beginning of astronomy. Atlantis was considered the cradle of the civilization, and while J.S. Bailly proposed that it was located in the Arctic and tried to convince Voltaire about his idea, G.R. Carli maintained the classical position in the Atlantic Ocean. Bailly's arguments and their criticism by Carli are of some interest, since they remind of the general problems that still plague cultural astronomy and archaeoastronomy. I will mention briefly the methodological problem of demonstration and evidence in these fields, and finally I will suggest a possible psychological limit when dealing with these topics.

## 1. Introduction

More than two centuries ago, some scholars discussed the myth of Atlantis in the context of the most recent studies in natural philosophy and human geography. The reason for those discussions was the interest in the problem of the beginning of civilization and, in particular, the role played by astronomy. Gian Rinaldo Carli (1720-1795) was one of those scholars. Born in Capodistria, a town of the Venetian Republic (today Koper, Slovenia), he had many cultural interests, particularly in political economy, and published a work on money and finance that the experts in the field consider of some relevance. He moved to Milan when the town enjoyed a significant economic and cultural progress, the so-called Milanese Enlightenment, under the Austrian rulers (Trampus 2006; Reinert 2012; Israel 2011). In 1765 Carli was appointed chairman of the governmental structure of public economy in Milan. After his retirement he published a book, *Le Lettere Americane* (*American Letters*), in 1780 with further editions until 1783. Probably, Carli took inspiration from other similar works, such as the *Lettres sur l'origine des sciences* and the *Lettres sur l'Atlantide de Platon* published by Jean-Sylvain Bailly (Paris, 1736-1793) just few years before. Bailly was a French astronomer and mathematician, who, after having published some scientific papers (see e.g. Delambre 1827, pp. 735-748), had abandoned the astronomical observations in order to focus on the history and mythology of astronomy. He had argued there must have been an 'antediluvian' culture which had excelled in astronomy, and for this thesis he became well known (Edelstein 2009). He gained also a high literary reputation and was a member of the Academy of Sciences, the French Academy and the Académie des Inscriptions. During the French Revolution he served as the first president of the National Assembly, he was elected mayor of Paris, but few years later he was guillotined (Smith 1954; see the biography by Lalande 1803, pp. 730-736)[1].

The purpose of the present note is to illustrate briefly the discussions that took place in those decades about America, Americans and Atlantis, as can be deduced from the *American Letters*, and to point out some connections with the present problems of cultural astronomy and archaeoastronomy. I do not mention all the hypothesis about the reality and the location of Atlantis that have been proposed from ancient times up to now, and I just note the plethora of books and papers (most would seem just pseudo-science) dedicated to this topic.

## 2. The *American Letters*

The book is neither a treaty nor an essay. It was composed rather freely, so the style is colloquial rather than academic. The book is divided in three volumes, and its content may be characterized by two main points: 1) a discussion of American pre-Columbian civilizations, and a detailed comparison of such civilizations with those in Europe, Africa (Egypt) and Asia; 2) a detailed discussion of Atlantis, in the context of the origin and evolution of Earth, and a comparison with the hypothesis of Bailly.

As regards the Pre-Columbian civilizations and the native Americans, the main references of Carli were the *Recherches philosophiques sur les Américains* of 1771 by DePauw, the *Journal du voyage fait par ordre du roi au equateur* of 1750-1751 by La Condamine, the *Histoire des deux Indes* of 1770-1780 by the Abbé Raynal, *The History of America* of 1777 by W. Robertson, and the *Storia antica del Messico (Historia Antigua de Mexico)* of 1780-1781 by F.J. Clavijero. As regards the origin of the Earth and Atlantis, the main references were *Les époques de la Nature* of 1778 and the *Histoire Naturelle, générale et particulière* by Buffon, the *Histoire de l'astronomie ancienne* of 1775 and the *Lettres sur l'Atlantide de Platon et sur l'ancienne histoire de l'Asie* of 1779 by Bailly. Carli, however, had also a deep knowledge of classical and modern writers, and included many quotations from them. One may note that the publication dates of the main bibliographical references are rather close in time, and this suggests that the debate about Americans was lively in those years (in the same period the United States got the independence).

Carli's work was appreciated by his contemporaries, it was translated, and a copy was sent by the Italian publisher also to Benjamin Franklin (Wolf and Hayes, 2006, p. 176; Miller, 1930).

## 3. Pre-columbian civilizations

In the first volume of the *American Letters* Carli criticized Cornelius De Pauw and his approach in the researches. DePauw (1739-1799) was a Dutch geographer and diplomat and published a study of American Indians in French; part of the work was translated into English. As remarked by Carli, DePauw had no hesitation in rejecting the observations of those with personal experience of the Americas, if they differed from his preconceptions. Despite this, he was reputed in his lifetime to be Europe's foremost authority on the Americas, although he never visited the continent. De Pauw was of the opinion that the American natives were inferior to European natives, and that this inferiority was partly due to American climate and geography. Carli, on the other hand, appreciated the work by Francisco Javier Clavijero (Francesco Saverio Clavigero, 1731-1787), a Jesuit born in Mexico who, after the expulsion of the congregation from that country in 1767, lived in Italy (e.g. Ronan 1977). Clavijero became aware of the extent of European ignorance about the nature and culture of pre-Columbian Americans after reading not only the essay of DePauw, but also those of Raynal and

---

[1] *"L'Académie des sciences devait l'adopter comme un de nos plus habiles astronomes, celle des inscriptions comme un de nos plus savans chronologistes, et l'Académie Française comme un de nos meilleurs écrivains"* (Lalande 1803, p. 733). *"Nommé successivement premier député de Paris aux états-généraux de 1789, premier président de l'assemblée qui se déclara constituante, et premier maire de Paris"* (Delambre 1827, p. 736).

Robertson (Clavigero, 1780, pp. 18-21; English trans. pp. V-VI, XXV-XXVI), and then he decided to write the history of Mexico[2].

Carli pointed out the strong similarities of American natives with other populations, as regards for example the weapons, tools and strategies in the wars; he remarked the valour of Mexicans (Aztecs) in their fights against Spaniards, and compared it with that of ancient Persians of king Darius. He illustrated the society of the Incas in Peru, their laws and lifestyles, commenting on the happiness of a theocratic monarchy that ruled a sort of communist society. He described in detail their religion, temples and customs, and the exact knowledge of astronomical phenomena. He suggested an analogy between *quipos* of Incas and ropes with knots of ancient China as a numerical system. Carli moreover said that Americans actually taught to Europeans some techniques, such as specific stone works and a better dyeing technique of fabrics.

In the second volume he compared the two hemispheres of the world, the one that included Europe, Asia and Africa (and Australia), and the other that included the Americas, and he stressed again on the analogies between American populations on the one hand, and Egyptians, ancient Europeans and Asians on the other, as regards customs and religion. He included also some fancy linguistic analogies, but he criticized the arbitrary use of etymologies done usually by scholars of his time. For instance, he compared the names Atlas and Atlantis with "*atlan*" of Aztec towns. He concluded that in a remote epoch there had been communications between the two hemispheres, across the Atlantic Ocean and Atlantis, on the one side, and across the North Pacific Ocean on the other. He recalled the tradition reported by the ancient writers of a large continent located to the West of Europe and Africa. He recalled also the tradition among the Aztecs regarding some persons who leaved from Mexico and went to a land located to the East; the emperor Montezuma thought of the Spaniards as the descendants of such people.

Carli compared the astronomical knowledge of Americans with that of ancient Egyptians and Greeks, and he remarked the accuracy of their calendars. Contrarily to what affirmed by DePauw, the Americans were not ignorant barbarians, or something less than men; they had a similar knowledge of astronomy to that of Greeks, and in some respects they were more advanced, as regards for example the small corrections to be applied for the motion of the Sun and the Moon.

## 4. Atlantis

From the analogy of customs, religion, writing system (hieroglyphs), ceremonies, terms and names, and astronomy, Carli concluded therefore that long time ago there were communications between the two hemispheres. He was certain of the historical truth of the traditional tales about Atlantis reported by the ancient writers. Therefore he drew the conclusion that there was a large island, a continent, in the Atlantic Ocean. He imagined that the present day islands of Atlantic Ocean are the remains of the highest mountains of Atlantis, whose extent should have been very approximately from about 40 degrees to -40 degrees of latitude. Since the native Americans did not know iron and money, such a continent should have disappeared before the discovery of those tools, in a remote epoch, several thousand years ago, and before the biblical flood.

He stressed on the need of a reliable chronology, and he remarked the importance of astronomical data for that application. From a detailed analysis of the cosmic cycles adopted in Egypt, Babylonia and China, Carli suggested that the epoch of Atlantis had been about 6600 years ago. Then he mentioned a possible catastrophic event that destroyed the island, the collision of a (large) comet with the Earth, which produced large effects. Before the collision, the Earth was slightly closer to the Sun and the year length was 360 days, the eccentricity was zero, and the inclination of Earth rotation axis on the ecliptic was different from today. The oceans were distributed differently, and the abrupt dynamical changes due to the collision produced a redistribution of the water. As a support, Carli mentioned the marine fossils found on the mountains, a demonstration that long time ago the water really covered them. Note that few years before, the astronomer Lalande (1773) published an essay on the comets approaching the Earth, that was misunderstood and caused incredible panic and hysteria in France: the people expected the end of the world. Two years later, another French astronomer, Du Sejour, published an essay with a more accurate estimate of the probability of the collision with a comet, 1/758730, to be compared with the estimate by Lalande, 1/76000. Carli compared the two values with the probability 1/11748 to win a terna at lotto, and commented that it happens, indeed.

---

[2] Clavigero (1807), after having listed the Spanish and Mexican writers on Mexico, wrote: "*we could add a long catalogue of French, English, Italian, Dutch, Flemish, and German writers, who have written either designedly, or accidentally, on the ancient history of that kingdom; but after having read many of them, to obtain assistance to this work, I found none who were of service except the two Italians, Gemelli and Boturini, who having been in Mexico, and procured from the Mexicans many of their paintings, and particular intelligence concerning their antiquity, have contributed, in some measure, to illustrate their history. All the others have either repeated what was already written by Spanish authors mentioned by us, or have altered facts, at their own discretion, to inveigh the more strongly against the Spaniards, as has lately been done by M. de Paw, in his Philosophical Enquiries concerning the Americans*". In an appendix to his second volume, Clavigero (1780) commented Carli's *Lettere Americane* that had been just published (first edition of 1780), with both some appreciations and some criticisms.

In summary, Carli gave some credit to the traditional Plato's account, and he remarked that such a tradition of Atlantis, which was present in Europe and Africa (Egypt), did not exist in Asian cultures. Atlas should have been the king of the large island of Atlantis located in the Atlantic Ocean; he invented astronomy, and diffused it on the East side, in Europe and Africa, and on the West side, in America. It was a sort of a (cultural) colonization of both hemispheres.

## 5. Buffon, Bailly and Voltaire

The supposed consequences of the collision with a comet were discussed by Carli also in the third volume, where he illustrated the hypothesis of Buffon about the evolution of the Earth, and criticized the idea of Bailly that Atlantis was located in a very northern region of the Arctic.

The Comte de Buffon (1707-1788) tried to estimate the evolutionary time of the Earth starting from its formation as a piece of hot solar material. He assumed a law of heat dissipation, even though the nature of heat was not yet clear at his time; the reasoning of Buffon was based approximately on the temperature as a measure of the amount of heat. A hot sphere (with slight depressions at the poles; Bailly 1777, p. 329) such as the Earth should have become cooler starting from the polar regions. At the end of 17th century, Olaus Rudbeck (1630-1702) wrote a treatise in Swedish where he purported to prove that Sweden was Atlantis, the cradle of civilization, and Swedish was the original language of Adam, from which the other languages had evolved. So it is not surprising that, taking Buffon's ideas into account, Bailly placed Atlantis not in Sweden but in the Arctic, in the Spitsbergen Island, that he considered the only habitable place in the initial very hot Earth. According to Bailly, the civilization then spread from Spitsbergen to Asia.

In 18$^{th}$ century the common idea was that Asia (India) was the true cradle of civilization, and Bailly tried to convince the old Voltaire (1694-1778) that astronomy and science had been invented in Atlantis-Spitsbergen and then spread to Asia, and hence the very ancient writings of India should have been just fables or myths and were not original astronomical knowledge. It is worth to recall the correspondence between the two scholars (Bailly 1779; pp. 1-9), since some questionable considerations by Bailly are still a problem in cultural astronomy. Voltaire replied to Bailly that he did not intend to debate with him, he wished just to learn since he was only an old blind man asking his help to find the way[3]. He asked why was it not believable that these Indian inventors (ancient brahmana) had invented also astronomy, in their good climate, since they actually needed more the astronomy to regulate the works and feasts than the fables to govern the men. Bailly commented that the need of things creates the necessity of inventing them, but that is not always possible. Even though the fables are based on some truth, they are the product of imagination; the scientific truths are the results of works, time and genius. The imagination is used before the advent of the kingdom of the reason: the adult that is involved in sciences does not come back to the plays of his childhood. It is because the Indians invented those fables that they were not able to invent and improve astronomy. One should note that Bailly maybe thought of an evolution of the populations, with a sort of children mentality at their beginning; in fact, the last sentence in his comment is: "*un palais n'est point bâti par des enfants*", a palace is not built at all by children. Voltaire noted that it is possible that an ancient population had instructed the Indians; but was it not permitted to doubt, in case there were no information about such an ancient people? Bailly replied that the doubt is always allowed in the sciences. However, the doubt must have some limits; not all the truths can be demonstrated as in the case of the mathematical truths. The mankind would have to lose too much if it had to reduce to this only class. The balanced testimonies, the weighted probabilities, the fables compared and clarified one with the other, form all together a strong light that can lead to evidence. Therefore, there are reasons to believe and not to doubt[4].

---
[3] It would seem that Voltaire wrote wittily using naively flattering expressions.

Bailly tried to be convincing with his eloquence, but he did not look persuasive. His position was weak, as noted by Carli, who used a picturesque analogy, mentioned in the next section. As already remarked, similar problems about what should be believable in our field of cultural astronomy are still with us. Carli criticized the procedure adopted by Bailly to combine pieces of ancient writings to support his opinion. He confessed he could not follow such a flood of fables, poetry fantasies, ambiguities of writers, of apocryphal writings and fragments. He said he was confused, and could not believe Bailly's conclusion that he had proven the origins in the Arctic. I note by the way that the above messy procedure is adopted also today by some would-be archaeoastronomers.

## 6. A romance by a magician?

Carli affirmed that he wrote in a colloquial way, and he did not pay attention to the perfection of the style: one had to take his thoughts just as simple hypotheses, and his *Lettere* had to be considered a similar 'poem' to *La pluralité des mondes* of Fontenelle (1657-1757). In the conclusion of the second volume, he remarked that he wrote a poem in prose just as a pastime.

On the other hand, according to Carli, Bailly was a terrible 'magician', or illusionist, since he was able to create a strong illusion in the reader, who could not resist. By means of an alluring violence, he leaded the mind of the reader to think and believe what Bailly wished he had to believe. However, Carli was convinced that Bailly, just as an illusionist, knew perfectly the truth, and he fooled the scholars into believing a different reality. Bailly, according to Carli, knew very well that all the traditions of ancient writers proved that Atlantis and Atlantideans were located west of Africa; but since he was engaged in proving that the sciences, and astronomy, had been conveyed to India and China from the mountains of Asia, he was therefore compelled to make the readers to believe that Atlantideans, the first astronomers, came also from northern Asia (Carli 1783, III, Letter VII, p. 84). As a confirmation of his opinion about the 'magician', Carli quoted the following statement of Bailly: "*si vous m'ordonnez de continuer les faits que je vous ai rapportés, de concilier Platon avec Plutarque et de vous faire un roman ... je puis vous obéir*" (Carli 1783, III, Letter IX, p. 135)[5]. That is, according to Carli, the previous reasoning about Atlantideans in Spitsbergen should be considered a romance, since he could not believe it was a scientific essay.

Maybe it is worth noting here that Bailly had acquired quite a reputation: although d'Alembert and Condorcet derided him as a mystical buffoon, in other academic, as well as more popular circles, he was hailed as a great historian and scientist, and in the final decade of the Old Regime

---

[4] "*Le doute est toujours permis dans les sciences, c'est la pierre de touche de la vérité. Cependant le doute doit avoir des bornes; toutes les vérités ne peuvent pas être démontrées comme les vérités mathématiques. Le genre humain aurait trop à perdre, s'il se réduisait à cette classe unique. Les témoignages balancés, les probabilités pesées, les fables rapprochées & éclairées les unes par les autres, forment par leur réunion une lumière forte qui peut conduire à l'évidence. Et lorsque la philosophie avec ces secours arrive à des résultats fondés sur la nature des choses & des hommes, on a des raisons de croire & non pas de douter*" (Bailly 1779; pp. 6-7, note to a letter by Voltaire).

[5] Actually Bailly wrote "*Si vous m'ordonnez de combiner les faits que je vous ai rapportés, de concilier Platon avec Plutarque, & de vous faire un Roman: quelque danger qu'il y ait de se livrer à l'imagination devant un grand poëte, dont le génie enfanta des romans touchans, ou des fables embellies par les graces, je me souviens toujours de votre indulgence: je puis vous obéir, car en vous écrivant, je n'écris qu'au philosophe; je vous parle, non pour vous amuser, mais pour m'instruire*" (Bailly 1779 ; Lettre 24, p. 464, or p. 430 in other reprinting).

Bailly thus found himself in a position of great institutional power and prestige (Edelstein 2009). His friend Lalande however complained that his last researches and discussions (the essays on the fables, published posthumously) were more curious than useful and did not contribute to the progress of astronomy, a science he would have been able to influence. In particular he remarked Bailly's fixed ideas: Lalande wrote that, in spite of his own efforts, he was not able to bring back him to what it seemed the truth[6].

**7. Criteria for a possible demonstration**

As quoted in Sect. 5, Bailly wrote that the doubt must have some limits, since not all the truths can be demonstrated as in the case of the mathematical truths. According to him, there are three useful criteria: the balanced testimonies, the weighted probabilities, the fables compared and clarified one with the other; they form all together a strong light that can lead to evidence. Apparently, this is a plausible and reasonable approach. Should we therefore accept that Atlantis was in Spitsbergen?

In my opinion, there are similar crucial problems in the field of cultural astronomy and archaeoastronomy. In particular, the problem of the intentionality of the astronomical orientations of ancient buildings: when is it possible to conclude that such an intentionality is evident? Is there a rigorous demonstration? The big problem of the rigorous proof in archaeoastronomy is a longstanding one, and what I can do here is to write just some comments. Schaefer (2006) has proposed four reasonable and plausible criteria: (a) statistical significance of the alignments, (b) the archaeological information that might bear on intention, (c) the ethnographic evidence concerning the desires and knowledge of the builders, and (d) the astronomical case for the utility of the claimed alignments[7]. I suggest there exists a sort of analogy with Bailly's criteria, that is, (b) and (d) could be put in relation with the "*témoignages balancés*", (a) with the "*probabilités pesées*", and (c) with the "*fables rapprochées & éclairées les unes par les autres*". We could say therefore with Bailly that Schaefer's four criteria "*forment par leur réunion une lumière forte qui peut conduire à l'évidence*"; but before that one should clarify the meaning of 'evidence' in our field[8]. The point is that the conclusions about the intentionality, in order to be believable, cannot be just the result of a demonstration in the sense of the hard sciences, even though scientific (astronomical, statistical) methods were rigorously applied. Intentionality by itself seems to be out of the realm of physical sciences. It is in this respect that the sentence by Bailly: "*toutes les vérités ne peuvent pas être démontrées comme les vérités mathématiques*" would be sound[9]. The intentionality is a topic for disciplines other than physics and astronomy, that is humanistic disciplines such as anthropology and archaeology, which have their own principles and methods: the last word is up to them. Intentionality has nothing to do with the physical principles and laws of the hard science. I think this ambiguity of archaeoastronomy is usually misleading people: since the rigorous and precise methods yield rigorous demonstrations in the field of hard sciences such as astronomy, it is assumed tacitly by the public that the same occurs in archaeoastronomy, that is, the rigorous archaeoastronomical methods can prove the intentionality. This assumption cannot be accepted, since in general that is not true. Even the expression 'very probable' or 'probable' often used by archaeoastronomers with reference to the intentionality should be used with a lot of care, unless it is supported by archaeological methods.

---

[6] "*Les idées de Bailly étaient fixées, son parti était pris; et malgré mes efforts, je ne pus le ramener à ce qui me semblait la verité*" (Lalande 1803, p. 733).

[7] The goal of the Author was to answer the key question of whether we can demonstrate that the claimed alignments were intentionally built into the structures. That is, intention must be proved. Without proof of intention, all we would have is a fun urban myth (Schaefer 2006).

[8] Somehow, one might take into consideration also the often quoted statement by W.K. Clifford: "*it is wrong always, everywhere, and for anyone, to believe anything upon insufficient evidence*" (Clifford 1877). But it is not easy. According to Kelly (2014) the concept of evidence is central to both epistemology and the philosophy of science; however, philosophers themselves have offered quite divergent theories of what sort of things are eligible to serve as evidence. Kelly recalls that the concept of evidence has often been called upon to fill a number of distinct roles, and some of them stand in some measure of tension with one another. He mentions the growing awareness (in twentieth century philosophy of science) "*of just how often eminent, fully-informed and* seemingly *rational scientists have disagreed in the history of science*", since the bearing of evidence on theory is mediated by factors that might vary between individuals. "*Bayesian holds that what it is reasonable for one to believe depends both on the evidence to which one is exposed as well as on one's prior probability distribution. According to the Bayesian then, two individuals who share exactly the same total evidence might differ in what it is reasonable for them to believe about some question in virtue of having started with different prior probability distributions*" (Kelly 2014). The Author remarks that different responses to a given body of evidence might be equally reasonable, so the bearing of a given body of evidence on a given theory could become a highly relativized matter. "*For this reason, the capacity of evidence to generate agreement among even impeccably rational individuals is in principle subject to significant limitations*" (Kelly 2014).

[9] The sense of Bailly's sentence is clear. However, as regards mathematics itself, it should be recalled that even here there are discussions about truth (e.g. Benacerraf 1973), and there are arguments that apparently are not yet or cannot be settled (undecidable problems, undecidable statements).

What are proof, demonstration and evidence in our field? In my opinion, there is no clear answer yet. Aveni (2006) tried to discuss specifically the problem of evidence and intentionality, pointing out the limits and possible flaws of the approach by Schaefer, seen from an anthropological and ethnological point of view. In particular he remarked that tools and methods of physical sciences cannot be adapted to such human sciences. Ruggles (2011), however, criticized such a strong statement, recalling for example that the *"scientific method has become carefully adhered to in rock-art studies"*[10].

As a practical conclusion, I would recall McCluskey. If archaeoastronomy is anything, it is not just astronomy, or archaeology, or anthropology, or history. Its practitioners should master the methods and questions of their own disciplines, and strive – as best they can – to understand the methods and questions of other disciplines (McCluskey 2007). I think that, in order to be sure about such an understanding and to avoid illusions such as those of Bailly, maybe it would be better to collaborate or at least interact directly with the scholars of the other disciplines. This seems, however, to be a long way[11].

## 8. Human psyche

In their studies about Atlantis, origin of civilization and astronomy, Bailly and Carli adopted a sort of selection, by pointing out the 'proofs' supporting their own ideas and overlooking a bit their weak points. This biased procedure is quite common even today, and hence many of the contemporary interpretations of Atlantis, with few exceptions, should be considered pseudohistory, or pseudoarchaeology. However, it seems to me that there is an important difference. Carli considered his work as a sort of a novel or a romance; he was well aware of the simply hypothetical nature of his study. He liked to discuss the pros and cons, as much as he could into detail, but he was aware that it was just a sort of a play. Today the situation looks different; several scholars appear very convinced about their own ideas, even though it would not be reasonably the case. It looks a curious behaviour, a bit foolish. I suspect this is another bad product of the separation of the disciplines. There is no more the 'republic of letters' of 18th century, where there was a fruitful exchange of ideas between disciplines, and perhaps it was easier to understand each other.

---

[10] On the other hand, he concluded that *"identifying robust methodologies for weighing together the different types of data with which the cultural astronomer is faced in different situations, so as to infer the 'best' interpretation, remains at once the most challenging and the most pressing issue facing our 'interdiscipline' in the future"* (Ruggles 2011).

[11] It is opportune to quote here a Nature editorial, *The University experiment; Universities must evolve if they are to survive*, which contains the remark that *"one perennial issue is the departmental structure that keeps researchers mentally and physically separated"* (Nature, 514, p. 287; 16 October 2014). Just as an example of the problems in a field close to ours, I would mention the section *"Science and Archaeology: a difficult liaison"* in the paper by Knapp and Manning (2016) on the issue of the Late Bronze Age collapse in the Eastern Mediterranean. The *"difficult liaison"* should be that between natural sciences, such as paleoclimatology, and archaeology. The Authors note the problem of the intractability of both archaeologists and scientists who embrace a predetermined position, so the *"outcome is often a selective filtering of data and related information and an unwillingness to contemplate or envisage a counter position. The realm of science and archaeology, or science in archaeology, is necessarily and by definition interdisciplinary […] It is difficult or impossible to be expert in all areas, and difficult to be even handed to all evidence and to judge and criticise it appropriately on its merits"* (Knapp, Manning, 2016; p. 101).

Therefore, in the last two centuries there had been room for bizarre thoughts about fascinating topics such as Atlantis, Egyptian pyramids, and so on. Maybe it would be worth to discuss about a sort of pathology. I would suggest to adapt a title of a famous essay by Sigmund Freud, and to consider, so to speak, the 'foolish' behaviour as a manifestation of the 'psychopathology of everyday culture'. Freud studied the 'psychopathology of everyday life'; he analyzed faulty actions, giving support to two "*recurring statements … namely, that the border-line between the nervous, normal, and abnormal states is indistinct, and that we are all slightly nervous*" (Freud, 1914; p. 337). It seems to me that everybody, more or less, is suffering of an analogous psychopathology, as shown by our falling in love with our own ideas (e.g. the fixed ideas of Bailly), even when they are not much sound. It appears a very common phenomenon not only in the field of cultural astronomy[12], where quite often an idea cannot be proved but also cannot be disproved rigorously. However, assuming it were sound, this point would require a thorough study with the help of a psychologist or psychiatrist (Fontana, Antonello 2016).

## 9. Conclusion

In the second half of 18th century Atlantis was considered the cradle of the civilization; J.S. Bailly was convinced that it was located in the Arctic, while G.R. Carli criticized such an idea. Bailly's arguments and their criticism by Carli remind of the problems still present in cultural astronomy and archaeoastronomy, concerning the methodology for a rigorous demonstration and what could be considered as evidence in these fields. I suggested a possible psychological limit; probably it would be worth to discriminate what could be cultural astronomy from what could be pathology of the culture, being well aware of the fact that it is not an easy job at all.

---

[12] At the most, in archaeoastronomy the risk would be having to do with some 'fun urban myth'; but, on the other hand, such a myth could be of some importance for a local touristic valorisation. However, in other fields the situation can be very tragic, as shown by some extreme cases of alternative medicine treatments.